\documentclass[aps,prl,twocolumn,showkey,square,numbers,amssymb,amsmath,nobibnotes,footinbib]{revtex4}
\usepackage{esint}
\usepackage{bm,bbm}
\usepackage{amsmath}
\usepackage{amsfonts}
\usepackage{amssymb}   
\usepackage{verbatim}
\usepackage{booktabs,array}
\usepackage{times}
\usepackage{graphicx,color}
\usepackage{tabularx}
\usepackage{dsfont}
\usepackage{titletoc} 
\usepackage{microtype}
\usepackage{tikz}
\usetikzlibrary{decorations.markings}
\usepackage{tkz-graph}
\usepackage[caption=false]{subfig}
\usepackage{blkarray}
\usepackage{hhline}
\def\mc#1{\multicolumn{1}{c|}{#1}}

\DeclareBoldMathCommand\boldlangle{\left\langle}
\DeclareBoldMathCommand\boldrangle{\right\rangle}

\def\Ord{\mathcal{O}}
\newcommand{\be}{\begin{equation}}
\newcommand{\ee}{\end{equation}}

\newcommand{\de}{\mathrm{d}}

\newcommand{\abs}[1]{\left| #1 \right|} 
\newcommand{\pd}[2]{\frac{\partial #1}{\partial #2}} 
 
\let\baraccent=\= 
\renewcommand{\=}[1]{\stackrel{#1}{=}} 

\newcommand{\Tr}{\mathrm{Tr\,}}
\newcommand{\Res}{\mathrm{Res}}

\newcommand{\Cov}{\mathrm{Cov}}
\newcommand{\avg}[1]{\left< #1 \right>} 
\newcommand{\T}{\mathcal{T}}
\newcommand{\rs}{\rho^{\star}}
\def\im{{\rm i}}
\newcommand{\kk}{\kappa}
\newcommand{\EE}{\mathcal{E}}
\newcommand{\Cc}{\mathcal{C}}

\def\lm{\lambda_-}
\def\lp{\lambda_+}
\def\lpm{\lambda_\pm}

\def\Ht{\tau_\mathrm{H}}
\def\Sm{\mathcal{S}}

\newcommand{\EXP}[1]{e^{#1}}     

\begin{document}


\title{Statistical distribution of the Wigner-Smith time-delay matrix moments for chaotic cavities}

\author{Fabio Deelan Cunden} 
\affiliation{School of Mathematics, University of Bristol, University Walk, Bristol BS8 1TW, England\\
Dipartimento di Matematica, Universit\`a di Bari, I-70125 Bari, Italy\\
Istituto Nazionale di Fisica Nucleare (INFN), Sezione di Bari, I-70126 Bari, Italy}

\begin{abstract} 
We derive the joint distribution of the moments $\Tr Q^{\kk}$ ($\kk\geq0$) of the Wigner-Smith matrix for a chaotic cavity supporting a large number of scattering channels $n$. This distribution turns out to be asymptotically Gaussian, and we compute explicitly averages and covariances. The results are in a compact form and have been verified numerically. The general methodology of proof and computations has a wide range of applications. 
\end{abstract}


\maketitle

\textit{Introduction - }  Transport phenomena at quantum scales offer a lot of theoretical and experimental challenges.  For quantum systems whose classical limit is chaotic a remarkable universality emerges in the scattering processes. A phenomenological description of these universal features is provided by random matrix theory (\textsf{RMT})~\cite{GuhMulWei1998,Akemann2011}. The scattering process of an electron at energy $E$ through a cavity is efficiently described by a $n\small{\times} n$ scattering matrix $\Sm(E)$, where $n$ is the number of scattering channels. If the cavity accommodates classical chaotic motion, the basic idea of \textsf{RMT} is to promote $\Sm(E)$ to a \emph{random} scattering matrix, the most inexpensive position being to take $\Sm(E)$ uniformly distributed in the group of unitary matrices~\cite{BarMello94}, restricted only by fundamental symmetries. This corresponds to physical realization of cavities with ideal coupling (completely transparent contacts). Although this might appear a mathematical oversimplification of the problem, these circular ensembles have proved fruitful to describe the universal features of quantum transport in chaotic cavities~\cite{Bee97,MelKum04}. It is worth mentioning that the \textsf{RMT} predictions are confirmed by  semiclassical methods whereby transport observables are translated as sums of probability amplitudes over classical trajectories~\cite{Gutzwiller90,Ric00,Sieber2002,Muller2004}.

Some aspects of the scattering processes can not be described by $\Sm(E)$ alone, but require some knowledge of its variation with respect to the energy. In these situations the essential features are captured by the Wigner-Smith time-delay matrix~\cite{Smi60}, $Q(E):=-\im\,\Sm(E)^\dagger\,\pd{\Sm(E)}{E}$ (with $\hbar=1$), whose eigenvalues $\tau_1,\cdots,\tau_n$ are referred to as the {\it proper time-delays}.  Insisting on the `equal \emph{a priori} probability' assumption for the whole energy dependent scattering matrix ensemble $\Sm(E)$, the joint distribution of the (positive) rescaled inverse time-delays $\lambda_k\equiv 1/(n\tau_k)$ is~\cite{BroFraBee97_99}~:
  \begin{equation}
    \label{eq:Brouwer1997}
    \mathcal{P}_{\beta}(\lambda_1,\dots,\lambda_n)
    \propto
    \prod_{i<j}|\lambda_i-\lambda_j|^\beta
    \prod_k \lambda_k^{\beta n/2}\EXP{-(\beta n/2)\lambda_k},
  \end{equation}
with a Dyson index $\beta$ fixed by the time-reversal and spin-rotation symmetries of the system 
(the times $\tau_k\geq0$ are measured in units of the Heisenberg time $\Ht=2\pi\hbar/\Delta$, where $\Delta$ 
is the mean level spacing).

The Wigner-Smith matrix is mostly known in investigations of the time delay of quantum scattering~\cite{Eis48,Wig55}. More generally, there exist physical observables that are entirely determined by its \emph{moments} $\Tr Q(E)^{\kappa}$. The Wigner time delay $(1/n)\Tr Q(E)$, a bona fide measure of the time spent by the electron  in the ballistic cavity, in open quantum systems is closely related to the density of states \mbox{$\nu(E)=(1/2\pi) \Tr Q(E)$}~\cite{Friedel52,Krein53,Lehmann95,Texier03}. Higher moments turn out to be crucial in the AC transport~\cite{ButtThomPre93,GopMelBut96,ButPol05}~, e.g. in the low frequency expansion ($\omega\to0$) of the AC dimensionless conductance \mbox{$G(\omega)=[-\im\omega\Tr Q+(1/2)\omega^2\Tr Q^2+...]$}.

Results on the Wigner time delay ($\Tr Q(E)^{\kappa}$ with $\kk=1$) abound. These include the cumulants for finite and large $n$ (see~\cite{Simm11_12,MezzadriCMP13,Sieber14} and references therein) and also the large deviation tails~\cite{FyoSom97,Savin01,TexierPRL13}~. A few results on the second moment  ($\kk=2$) can be also found in the cases $n=2$~\cite{Pedersen98} and $n\gg1$~\cite{Grabsch14}~. 

A statistical description of the \emph{full} family of moments $\Tr Q(E)^{\kappa}$ ($\kk\geq1$) is still unavailable.
The aim of this Rapid Communication is twofold. First, to provide a simple proof that the moments of the Wigner-Smith matrix of a chaotic system are \emph{jointly} asymptotically Gaussian. Second, and more ambitiously, to compute the mean and the covariances of $\Tr Q(E)^{\kappa}$  \emph{for all} $\kappa$ at leading order in the number $n$ of  scattering channels. We will do so from the following angles: simplicity of the results, numerical efficiency and extendibility to other problems. 

Our proof of the large $n$ statistical behavior relies on standard statistical mechanical arguments recently evolved into rigorous results in the theory of large deviations.  About the averages and covariances we provide close and simple formulae. By controlling the behavior of the moments, one could easily derive the statistical distribution of any (analytic) function of the proper time-delays. In the derivation we will introduce a prescription to handle a covariance formula for random matrix ensembles~\cite{Ambjorn90,BrezinZee93,Beenakker94,Kanz96,CundenVivo}. We are confident that this prescription will turn out to be precious in many other problems.

\textit{Results - } For notational convenience we denote by \mbox{$\T_{\kk}:=n^{\kk}\Tr Q^{\kk}$} the rescaled moment  of the Wigner-Smith matrix of a chaotic system. To begin with, several elementary observations. The moments can be written as linear statistics $\T_{\kk}=\sum_{i}1/\lambda_i^{\kk}$, i.e. \emph{sum functions} of the inverse time-delays $\lambda_i$. Therefore, they admit an integral representation $\T_{\kk}=n\int\de\lambda\,\rho_n(\lambda)\lambda^{-\kk}$ in terms of the spectral density $\rho_n(\lambda)=n^{-1}\sum_{i}\delta(\lambda-\lambda_i)$. This representation will be convenient to investigate the setting of large number $n\gg1$ of scattering channels. The trivial case $\kk=0$ corresponds to the nonrandom quantity $\T_0=\Tr I_n$ (with $I_n$ the $n\small{\times}n$ identity matrix). 

In this work, we claim that for any $\beta>0$ the $\T_{\kk}$'s are \emph{jointly} asymptotically Gaussian. Hence, the averages $\avg{\T_{\kk}}$ and the covariance structure $\Cov(\T_{\kk},\T_{\ell})=\avg{\T_{\kk}\T_{\ell}}-\avg{\T_{\kk}}\avg{\T_{\ell}}$ are sufficient statistics for this family to leading order in $n$ (hereafter the angle brackets stand for the averaging with respect to \eqref{eq:Brouwer1997}). Here we compute exactly the leading order of $\avg{\T_{\kk}}$ and $\Cov(\T_{\kk},\T_{\ell})$ for \emph{all} $\kk,\ell\geq0$.

Throughout the paper, a prominent role will be played by the complex function 
\be
\Psi(z)=\frac{1}{2 z}\sqrt{(z-\lm)(z-\lp)}\ ,\quad\lpm=3\pm2\sqrt{2}\ . \label{eq:f}
\ee
In the derivation, it will be clear that $\Psi(z)$ is a relative of the limit density of the inverse time-delays (see \eqref{eq:rhostar} below) which by itself encodes both averages and fluctuations of random Wigner-Smith matrices.

The results can be summarized as follow. We find that the average $\T_{\kk}$ scales linearly with the total number of channels $n$, while the covariances are of order $\Ord(1)$
\be
\label{eq:avg_cov}
\avg{\T_{\kk}}\sim n r_{\kk}\qquad \Cov(\T_{\kk},\T_{\ell})\sim (1/\beta)\,\Cc_{\kk,\ell}\ .
\ee
We have computed the generating functions of the sequences of numbers $\{r_{\kk}\}$ and $\{\Cc_{\kk,\ell}\}$. They read explicitly 
\begin{align}
\label{eq:R}
R(z)&=\sum_{\kk\geq0}r_{\kk}z^{\kk}=\frac{1}{2}(3-z)- z\Psi(z)\\
\label{eq:G}
G(z,\zeta)&=\sum_{\kk,\ell\geq0}\Cc_{\kk,\ell}z^{\kk}\zeta^{\ell}=\frac{z}{\Psi(\zeta)}\frac{\partial\,}{\partial z} \frac{z\Psi(z)-\zeta\Psi(\zeta)}{\zeta-z}\ .
\end{align}

 In Table~\ref{tab:1}, we report the first values of  $\avg{\T_{\kk}}n^{-1}$ and $\Cov(\T_{\kk},\T_{\ell})$. They can be effortlessly generated as series coefficients of \eqref{eq:R}-\eqref{eq:G}.  As already discovered in~\cite{BerkKuip10,Novaes11}, the numbers $r_{\kk}=1,1,2,6,22,\dots$ ($\kk\geq0$) enumerate some combinatorial objects (see~\cite{Gou-BeauVauq88}) and they are referred to as the sequence of the `large Schr\"oder numbers'~\cite{Schroder}~.  We point out that the generating function \eqref{eq:R} slightly differs from the previously known expression~\cite{BerkKuip10}~. Our expression includes also the trivial case $ \kk=0$, that is $r_0=1$, a value compatible with the combinatorial interpretation of the large Schr\"oder numbers. The $\Cc_{\kk,\ell}$'s  in \eqref{eq:avg_cov}-\eqref{eq:G} are instead a new result and a combinatorial interpretation of them is elusive. One may verify by inspection that $G(z,\zeta)=G(\zeta,z)$ and therefore $\Cc_{\kk,\ell}=\Cc_{\ell,\kk}$. 
\begin{table}
\[\biggl[
\begin{array}{r}
\kk\quad:\\
r_{\kk}\quad:
\end{array}
\begin{array}{ccccccccccc}
$0$    & $1$ & $2$ & $3$ & $4$ & $5$  & $6$ & $7$ & $8$ & $9$ & \cdots \\
$1$& $1$ & $2$ & $6$ & $22$ & $90$ & $394$ & $1\,806$ & $8\,558$ & $41\,586$ & \cdots
\end{array}\biggr]
\]

\[
\Cov(\T_{\kk},\T_{\ell})\sim\frac{1}{\beta}\left[
\begin{array}{cccccc}
$0$ &  $0$ & $0$  &  $0$  &   $0$  &\cdots\\ \cline{2-3}
\mc{$0$} & $ 4 $ & \mc{$24$} & $132$ & $720$ & \cdots  \\
\mc{$0$} &  $24$ & \mc{$160$} & $936$ & $5\,312$   \\  \cline{2-3}
$0$ &  $132$ & $936$ & $5\,700$ & $33\,264$   \\
 $0$ & $720$ & $5\,312$ & $33\,264$ & $198\,144$  \\
\vdots & \vdots &           &     &  & \ddots           
\end{array}
\right]
\]
\caption{Top: leading order of $\avg{\T_{\kk}}n^{-1}$. This sequence is generated as $r_{\kk}=(1/\kk!)\partial_z^{\kk} R(z)|_{z=0}$ with generating function $R(z)$ in \eqref{eq:R}. Bottom: a block ($0\leq\kk,\ell\leq4$) of the covariance matrix  \mbox{$\Cov(\T_{\kk},\T_{\ell})$} at leading order in $n$ from \eqref{eq:avg_cov}. The entries are computed from the power expansion of $G(z,\zeta)$ in \eqref{eq:G}. We have emphasized  the $2\times2$ previously known sub-block~\cite{Grabsch14}~($\kk,\ell=1,2$).}
\label{tab:1}
\end{table}

\textit{Derivation - } Our derivation relies on some old and new results on \emph{linear statistics} $\sum_i f(\lambda_i)$ on random matrices. As already disclosed, the moments $\T_{\kk}=\sum_i f(\lambda_i)$ are linear statistics on the inverse time-delay matrix with $f(x)=x^{-\kk}$. We will first show that the family of moments $\{\T_{\kk}\}$ ($\kk\geq1$) is asymptotically Gaussian with averages that scale linearly with $n$ and covariances of order $\Ord(1)$. To show this, we resort to the Coulomb gas picture suggested by Dyson~\cite{Dyson62}. 
Using the spectral density $\rho_n(\lambda)=n^{-1}\sum_{i}\delta(\lambda-\lambda_i)$, for large $n$, the joint distribution \eqref{eq:Brouwer1997} of the $n$ inverse time-delays can be cast in the Gibbs-Boltzmann form  \mbox{$\mathcal{P}_\beta(\lambda_1,\dots,\lambda_n)\approx \EXP{-\beta n^2\EE[\rho_n]+o(n^2)}$}. The functional
\be
\EE[\sigma]=-\frac{1}{2}\iint\hspace{-1mm}\de\sigma(\lambda)\de\sigma(\mu)\,\log\abs{\lambda-\mu}+\int\hspace{-1mm}\de\sigma(\lambda)\,V(\lambda)  , \label{eq:functional}
\ee
is the energy of a 2D Coulomb gas (logarithmic repulsion) on the half-positive line ($\lambda_i=1/\tau_i\geq0$) in equilibrium at inverse temperature $\beta>0$  in a confining single-particle potential \mbox{$V(\lambda)=(\lambda-\log\lambda)/2$}.
It is known that for large $n$ the average spectral density has a  nonrandom limit $\rs(\lambda)=\lim_n\avg{\rho_n(\lambda)}$. Indeed, going back to the 2D Coulomb gas picture, the external potential $V(\lambda)$ is convex and sufficiently deep to guarantee the electrostatic stability of the 2D Coulomb gas. For large $n$, the density $\rho_n(\lambda)$ of the gas reaches a stable equilibrium configuration described by $\rs(\lambda)$. This equilibrium density is the (unique) minimizer of the energy functional~\cite{Johansson98} 
\be
\rs=\arg\!\min\left\{\EE[\sigma]\,\colon \sigma\geq0\,\,\text{and}\int_{\lambda\geq0}\hspace{-3.5mm}\de\sigma(\lambda)=1\right\}\ . \label{eq:rhostar}
\ee
Performing the minimization is a standard procedure~\cite{Tri57}, and one finds the limiting density of the 2D Coulomb gas (i.e. the inverse delay-times) supported on $\lambda\in[\lm,\lp]$~:
\be
\rs(\lambda)=\frac{\sqrt{(\lambda-\lm)(\lp-\lambda)}}{2\pi \lambda}\ ,\quad\lpm=3\pm2\sqrt{2}\ . \label{eq:rhostar}
\ee
In \textsf{RMT}, \eqref{eq:rhostar} is known as Mar\v{c}enko-Pastur distribution with concentration parameter $c=2$~\cite{MP}. Moreover, the spectral density  $\rho_n(\lambda)$ is self-averaging, meaning that the equilibrium density $\rs(\lambda)$ is the \emph{typical} configuration of the eigenvalues. More precisely, the spectral density $\rho_n(\lambda)$ satisfies a `large deviation principle' with \emph{speed} $\beta n^2$ and \emph{rate function} $ \Delta\EE[\rho_n]:=\EE[\rho_n]-\EE[\rs]$~\cite{Guionnet}~. The informal translation of this statement is that, for large $n$, the probability of a configuration $\rho_n(\lambda)$ of the Coulomb gas scales as $\Pr(\rho_n)\approx\EXP{-\beta n^2 \Delta\EE[\rho_n]}$~: the probability is exponentially depressed in $\beta n^2$ and the precise measure of
unlikeliness of an out-of-equilibrium configuration is given by the energy penalty with respect to the equilibrium configuration.  From this large deviation principle it is immediate to derive the asymptotic Gaussian behavior of the $\T_{\kk}$'s. Indeed, since small variations of $\rho_n(\lambda)$ correspond to small changes of the moments \mbox{$\T_{\kk}=n\int\de\lambda\,\rho_n(\lambda)\lambda^{-\kk}$}, a saddle-point approximation gives $\Pr\left(\T_{\kk}=nt\right)\approx \EXP{-\beta n^2 \psi(t)}$ with
\be
\label{eq:ratefucnt_contr}
\qquad \psi(t)=\min\{\Delta\EE[\sigma]\,\colon \int\de\lambda\,\sigma(\lambda)\lambda^{-\kk}=t \}\ .
\ee
 In words, the probability that $\T_{\kk}$ takes value around a given value $nt$ is driven by the most probable (less energetic) configuration of the gas compatible with the prescribed value of  $\T_{\kk}$. In a more formal way, by the \emph{contraction principle} (see for instance~\cite{Touchette09}), the random variable $n^{-1}\T_{\kk}$ being a functional of $\rho_n(\lambda)$  satisfies a large deviation principle with \emph{same speed} $\beta n^2$ and rate function given by \eqref{eq:ratefucnt_contr}. One can extend the same reasoning to the joint distribution of $(\T_{\kk_1},\T_{\kk_2},\dots,\T_{\kk_v})$ and conclude that $\Pr\left(\T_{\kk_1}=nt_1,\dots,\T_{\kk_v}=nt_v\right)\approx \EXP{-\beta n^2 \psi(t_1,\dots,t_v)}$ with  rate function $\psi(t_1,\dots,t_v)=\min\{\Delta\EE[\sigma]\,\colon\int\de\lambda\,\sigma(\lambda)\lambda^{-\kk_i}=t_i\ ,\,i=1,2,\dots,v \}$. From this asymptotic behaviour one computes the large $n$ cumulant generating function~\footnote{The large deviation principle justifies this saddle-point computation. This is an instance of a more general result known as Varadhan's lemma. See~\cite{Touchette09}.} $J(s_1,s_2,\dots,s_v)=\lim_{n} (1/n^2)\log\avg{\mathrm{exp}[-\beta n\sum_{i=1}^v s_i\T_{\kk_i}]}$, from which one may extract the joint cumulants at leading order in $n$ through the formula  $C(\T_{\kk_1},\T_{\kk_2},\dots,\T_{\kk_v})=
n^2(-1/\beta n)^{m_1+\cdots+m_v} \partial_{s_1}^{m_1}\partial_{s_2}^{m_2}\cdots\partial_{s_v}^{m_v}J(0,0,\dots,0)
$. 
Therefore, the mixed cumulants behave for large $n$ as $C(\T_{\kk_1}^{m_1},\T_{\kk_2}^{m_2},\dots,\T_{\kk_v}^{m_v})=\Ord(n^{2-(m_1+m_2+\cdots +m_v)})$, for $m_i,\kk_i\geq1$. This shows that the averages grow like $n$, the covariances stay bounded in $n$, and the mixed cumulants of order $\geq 3$ decay faster than the second order ones. Below we will compute the covariance structure  showing explicitly that it is non-trivial. This argument proves that  $(\T_{\kk_1},\T_{\kk_2},\dots,\T_{\kk_v})$ are jointly asymptotically Gaussian. Then, any finite dimensional distribution of the family $\{\T_{\kk}\}$ ($\kk\geq1$) is asymptotically Gaussian and this concludes the proof.

\begin{figure}[t]
\centering
\includegraphics[width=1\columnwidth]{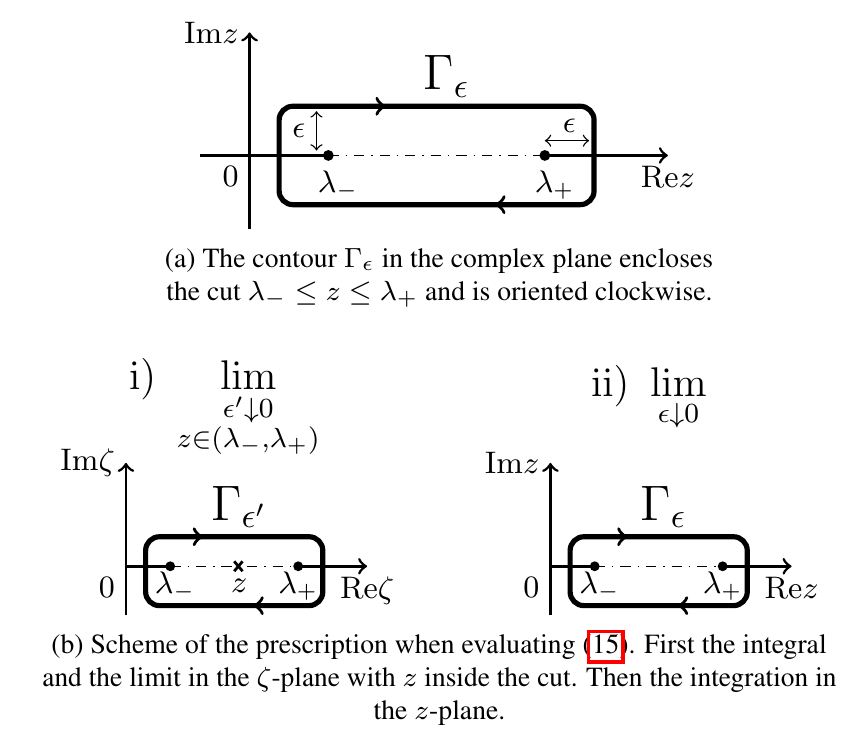}
\caption{Contours of integration used to evaluate \eqref{eq:int_an} and \eqref{eq:tech2}.} 
\label{fig:contour1}
\end{figure}

The asymptotics of the average moments $\avg{\T_{\kk}}$ is linearly related to the limiting spectral density \eqref{eq:rhostar} by  $\avg{\T_{\kk}}\sim n\int\de\lambda\,\rs(\lambda)\lambda^{-\kk}$ (recall that $\rs(\lambda)$ is identically zero outside $[\lm,\lp]$) . A standard way to evaluate the integral is the following. By using the complex function $\Psi(z)$ defined in \eqref{eq:f}  let us consider the integral in the complex plane  
\mbox{$I_{\epsilon}=\int_{\Gamma_{\epsilon}}\de z\,\Psi(z)f(z)$},
where the path of integration $\Gamma_{\epsilon}$ (see Fig.~\ref{fig:contour1}) in the complex plane is a clockwise oriented contour which encloses the interval (the cut) $[\lm,\lp]$ and $f(z)$ is analytic in a neighborhood of the cut. By inspection 
\be
\lim_{\epsilon\downarrow0}\int_{\Gamma_{\epsilon}}\de z\,\Psi(z)f(z)=2\pi\im\int\de\lambda\,\rs(\lambda)f(\lambda)\ . \label{eq:int_an}
\ee
On the other hand, if $\epsilon$ is sufficiently small, $I_{\epsilon}=-2\pi\im\sum_{z_k}\epsilon_{z_k}\Res\left(\Psi(z)f(z);z_k\right)$, where the sum of the residues is over all the poles $z_k$'s in the extended complex plane (in this case $z_k=0$ and the point at infinity), with the appropriate sign $\epsilon_{z_k}=\pm1$. Hence, with $f(\lambda)=\lambda^{-\kk}$ we get
\be
\label{eq:avgder2}
\avg{\T_{\kk}}\sim n\left[\Res\left(z^{-\kk}\Psi(z);\infty\right)-\Res\left(z^{-\kk}\Psi(z);0\right)\right].
\ee
The residues of the function $\Psi(z)z^{-\kappa}$ are trivially related to the power series of $\Psi(z)$ itself. Moreover, one can check that the residue at $z=\infty$ is nonzero only for $\kappa=0,1$.   Then, by introducing the Laurent series expansion of $\Psi(z)=\sum_{\kk=-1}^{\infty}a_{\kk}z^{\kk}$ around $z=0$ we evaluate  \eqref{eq:avgder2} as $n\left[\delta_{0\kk}+\delta_{1\kk}+(1-\delta_{1\kk})(1-\delta_{0\kk})a_{\kk}\right]$,
where $\delta_{ij}$ is the Kronecker delta. It is now a matter of elementary manipulations to recover the generating function \eqref{eq:R}.

We now turn to the computation of the covariance structure $\Cov(\T_{\kk},\T_{\ell})$ at leading order in $n$. Starting from scratch we have
\begin{align}
\Cov(\T_{\kk},\T_{\ell})&=n^2\hspace{-1mm}\iint\hspace{-1mm}\de\lambda\de\lambda'\Cov\left(\rho_n(\lambda),\rho_n(\lambda')\right)\left[\frac{1}{\lambda}\right]^{\kk}\hspace{-1mm}\left[\frac{1}{\lambda'}\right]^{\ell}\nonumber\\
&\sim- \iint\hspace{-1mm}\de\lambda\de\lambda'\,\mathcal{K}(\lambda,\lambda')(1/\lambda)^{\kk}(1/\lambda')^{\ell}
\end{align}
where in the first line we have used the fact that the only randomness is in the inverse time-delays and we have introduced the (smoothed) two-point  kernel $\mathcal{K}(\lambda,\lambda')=-\lim_n n^2\Cov\left(\rho_n(\lambda),\rho_n(\lambda')\right)$. 
In our setting the kernel can be computed exactly~\cite{Ambjorn90,BrezinZee93,Beenakker94,Kanz96,CundenVivo} and we have
\be
\label{eq:cov1}
\Cov(\T_{\kk},\T_{\ell})\sim\frac{1}{\beta\pi^2}\,\int\hspace{-2mm}\fint_{\lambda_-}^{\lambda_+}\!\!\!\!\!\ \de\lambda\de\lambda'\, \Theta(\lambda,\lambda')\frac{(1/\lambda')^{\ell}}{\lambda'-\lambda}\frac{\de (1/\lambda)^{\kk}}{\de\lambda}\ ,
\ee
where  $\fint$ denotes the Cauchy principal value and $\Theta(\lambda,\lambda')=\sqrt{((\lp-\lambda)(\lambda-\lm))/((\lp-\lambda')(\lambda'-\lm))}$. Remarkably, the two-point kernel (and hence the covariance structure) depends \emph{only} on the edges $\lpm$ of the limiting spectral density $\rs(\lambda)$, one of the manifestations of the universality in \textsf{RMT}. Solving \eqref{eq:cov1} is the main technical issue we managed to overcome. 
In the evaluation of \eqref{eq:cov1} we will use again the complex plane, but it should be borne in mind that this is merely a
device for calculating (there is nothing intrinsically complex about the Cauchy principal value). Expressing $\Theta(\lambda,\lambda')$ in terms of $\rs(\lambda),\rs(\lambda')$, and using similar arguments as before, we obtain the following identity for the integral in \eqref{eq:cov1}
\begin{align}
\int_{\lm}^{\lp}&\!\!\!\de\lambda\,\biggl\{-\kk\frac{\rs(\lambda)}{\lambda^{\kk}}\fint_{\lambda_-}^{\lambda_+}\!\! \frac{\de\lambda'}{\lambda'-\lambda}\frac{\,\,(1/\lambda')^{\ell+1}}{\rs(\lambda')}\biggr\} \nonumber\\
\label{eq:tech2}
=-\frac{\kk}{4\pi^2}\lim_{\epsilon\downarrow0}&\int_{\Gamma_{\epsilon}}\!\!\!\!\de z\,\biggl\{\frac{\Psi(z)}{z^{\kk}}\left[\lim_{\epsilon'\downarrow0}\int_{\Gamma_{\epsilon'}}\! \frac{\de\zeta}{\zeta-z}\frac{\zeta^{-\ell-1}}{\Psi(\zeta)}\right]\biggr\}\ .
\end{align}
Here, $\Gamma_{\epsilon},\Gamma_{\epsilon'}$ are again two contours enclosing the cut $[\lm,\lp]$ as in Fig.~\ref{fig:contour1}. 
The main usefulness of
this strategy is evident: relocating to the complex plane, the principal value integral becomes irrelevant as long as we adopt the following prescription. One should \emph{first} evaluate the integral in $\zeta$ and the limit of small $\epsilon'$ with fixed $z$ \emph{inside the cut} $[\lm,\lp]$. This way, the singularity $\zeta=z$ becomes immaterial. It turns out that the limit in the square bracket is a function of $z$ analytic in a neighborhood of the cut. Then, one can treat the integral in $z$ on a sufficiently small contour $\Gamma_{\epsilon}$ with no worries. This strategy is schematically summarized in Fig.~\ref{fig:contour1}. Using this prescription one can easily rewrite \eqref{eq:tech2} as
\begin{align}
\label{eq:avg2}
\kk\,\Res\left(\frac{\Psi(z)}{z^{\kk}}{\small\times}
\Res\left(\frac{\zeta^{-(\ell+1)}}{\Psi(\zeta)(\zeta-z)};\zeta=0\right)
;z=0\right).
\end{align}
The residue in $\zeta=0$ is given by $d_{\ell}(z)=-z^{-\ell}\sum_{j=0}^{\ell-1}d_jz^j$, where, surprisingly, the $d_j$'s have a combinatorial interpretation as the `central Delannoy numbers' (Sloane's A001850)~\cite{Sloane,Sulanke}~. At this stage we end up with the function $\Psi(z)d_{\ell}(z)/z^{-\kk}$ analytic in a neighborhood of the cut as anticipated. The calculation of the residue in $z=0$ is again related to the Laurent expansion of $\Psi(z)$ around $z=0$. At this level it is evident that all the steps are invariant under the simultaneous exchanges $\kk\leftrightarrow\ell$ and $z\leftrightarrow\zeta$. Carrying out a rearrangement of the terms we get the final result \eqref{eq:avg_cov}-\eqref{eq:G}.
\begin{figure}[t]
\centering
\includegraphics[width=1\columnwidth]{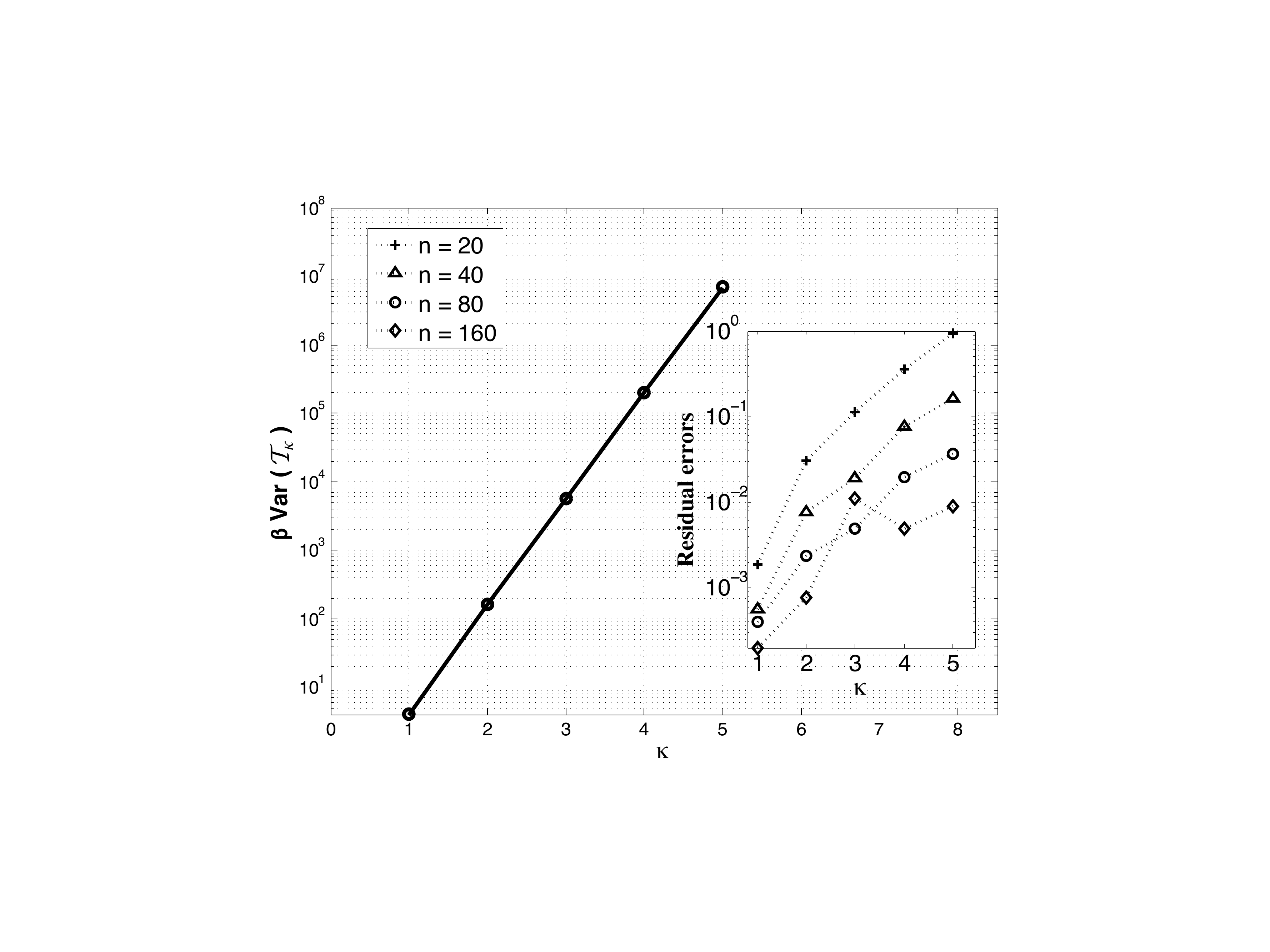}
\caption{Variances of $\T_{\kk}$ for $\kk=1,...,5$. The data points are obtained from numerical simulation of complex ($\beta=2$) Wishart-Laguerre matrices ($n=80$ in the main figure). The solid line connects the theoretical prediction in \eqref{eq:avg_cov}-\eqref{eq:G} with $\kk=\ell$. In the inset we demonstrate that the residual errors $\abs{(\sigma^2_{\mathrm{num}}-\sigma^2_{\mathrm{th}})/\sigma^2_{\mathrm{th}} }$ decrease for increasing values of $n$.}
\label{fig:numerical}
\end{figure}

\textit{Numerical Simulations - } The measure \eqref{eq:Brouwer1997} is known as the Wishart-Laguerre ensemble in \textsf{RMT}, and also corresponds to the distribution of the eigenvalues of Wishart matrices $\mathcal{X}^\dagger\mathcal{X}$, where  $\mathcal{X}$ is a $n\small{\times}(2n-1+2/\beta)$ matrix with independent standard Gaussian entries. Using an efficient sampling based on the tridiagonal construction \cite{Dumitriu02}, we have estimated $\avg{\Tr Q^{\kappa}}$ and 
$\Cov(\Tr Q^{\kk},\Tr Q^{\ell})$ from a sample of about $N=10^7$ spectra of complex ($\beta=2$) Wishart matrices for several sizes $n$. The agreement with the prediction \eqref{eq:avg_cov} is a further demonstration of  the correctness of our computations. Our findings are summarized in Fig.~\ref{fig:numerical}. We also performed a careful analysis of the residual errors for increasing sizes of $n$. The behavior of these deviations ($\lesssim 1 \%$ for $n=20$ and $\lesssim 0.01 \%$ for $n=160$) is compatible with an expansion of the cumulants of $\T_{\kk}$ in powers of $1/n$  whose leading terms are \eqref{eq:avg_cov}.

\textit{Conclusions -}  We provided a so far unavailable description of the asymptotic joint statistics of all moments $\T_{\kk}=n^{\kk}\Tr Q^{\kk}$ of the Wigner-Smith matrix of chaotic cavities with ideal coupling.  Beside solving an intriguing problem in mesoscopic physics, we believe that the work could be extended in several ways. 

The technique employed to show the asymptotic normality and compute averages and covariance is tailored for a large class of matrix models (e.g. unitarily invariant ensembles) and can be easily applied to many physical problems with an underlying random matrix description.  In particular, the method presented here has already proved useful elsewhere: it has provided a closed-form solution for a  problem arising in the context of a general class of matrix models~\cite{Cunden15} (some of these models describe the planar approximation in field theories with large internal symmetry group).

It is worth mentioning that there exist several techniques to verify a central limit theorem for spectral linear statistics on random matrices. Perhaps some of these methods may be of use to compute the finite $n$ corrections to~\eqref{eq:avg_cov}. For results and methods on the Wishart-Laguerre ensemble see~\cite{MP2011}. 

We have seen that the covariance structure $\Cov(\T_{\kk},\T_{\ell})$  is determined by the numbers $\Cc_{\kk,\ell}$. It would be surely of great interest to understand the combinatorial meaning of this sequence. 
The results presented in this work may also have some challenging theoretical implications:  we offer a new family of \textsf{RMT} predictions whose verification by semiclassical methods would be a new strong evidence of the intimate connection between random matrices and quantum chaology. 

\textit{Acknowledgments - }  The author gratefully appreciated stimulating discussions with M.~Sieber. He is obliged to F.~Mezzadri, P.~Facchi and P.~Vivo for encouragement and helpful remarks on the manuscript. The author is also grateful to two anonymous referees for valuable remarks and  bibliographical comments. This work was supported by EPSRC Grant number EP/L010305/1 and partially supported by Gruppo Nazionale di Fisica Matematica GNFM-INdAM.

\end{document}